\documentclass[hyperref={pdfstartview={FitBH -32768},
                         pdfpagemode=FullScreen,
                         plainpages=false,
                         colorlinks=true,11pt}
              ]{article} 

\usepackage[utf8]{inputenc} 
\usepackage{hyperref}
\usepackage{latexsym}
\usepackage{amsfonts}

\usepackage{geometry} 
\usepackage{graphicx} 
\usepackage{color}
\usepackage{crayola}

\usepackage{booktabs} 
\usepackage{array} 
\usepackage{paralist} 
\usepackage{verbatim} 
\usepackage{subfig} 

\usepackage{fancyhdr} 
\usepackage{sectsty}
\allsectionsfont{\sffamily\mdseries\upshape} 

\usepackage[nottoc,notlof,notlot]{tocbibind} 
\usepackage[titles,subfigure]{tocloft} 

\newcommand{\clock}{\count254=\time \divide\count254 by 60
 \count255=\count254 \multiply\count255 by -60
 \advance\count255 by \time
 \ifnum\count254<10 0\fi\number\count254\,:\,%
 \ifnum\count255<10 0\fi\number\count255}

\title{Drilling into Erasmus learning mobility flows\\ between countries  2014-2024}
\author{Vladimir Batagelj\\
IMFM Ljubljana,  UP IAM Koper, UL FMF Ljubljana\\
ORCID: 0000-0002-0240-9446\\
e-mail: \texttt{vladimir.batagelj@fmf.uni-lj.si}}
\date{} 

\newcommand{\RR}{\Bbb{R}}
\newcommand{\Net}{\network{N}}
\newcommand{\network}[1]{\mathcal{#1}}
\newcommand{\graph}[1]{\mathcal{#1}}
\newcommand{\vertices}[1]{\mathcal{#1}}
\newcommand{\arcs}[1]{\mathcal{#1}}
\newcommand{\Nodes}{\vertices{V}}
\newcommand{\Links}{\arcs{L}}
\newcommand{\edges}[1]{\mathcal{#1}}
\newcommand{\functions}[1]{\mathcal{#1}}
\newcommand{\cling}{\mathbf{C}}
\newcommand{\keyw}[1]{\textcolor{red}{\emph{#1}}}
\newcommand{\PFnet}{\mathop{\rm PFnet}\nolimits}
\newcommand{\Mw}[1]{\mathop{\raisebox{-1.5pt}{\mbox{\raisebox{2.5pt}{\fbox{\tiny $#1$}}\kern0.1pt}}}}

\graphicspath{{./pics/}{C:/Users/vlado/DL/data/erasmus/flows/}}

\begin{document}
\maketitle
\date{}

\begin{abstract}
Analyzing the Erasmus mobility network, we illustrate typical problems and approaches in analyzing weighted networks.

We propose alternative exploratory views on the network “Erasmus+ learning mobility flows since 2014”. The network has 35 nodes (countries), is very dense, and the range of link weights (number of visits) is very large (from 1 to 217003). An increasing transformation is used to reduce the range. The traditional graph-based visualization is unreadable. 

To gain insight into the structure of a dense network, it can be reduced to a skeleton by removing less essential links and/or nodes. We have determined the 1-neighbors and 2-neighbors subnetworks. The 1-neighbors skeleton highlights Spain as the main attractor in the network. The 2-neighbors skeleton shows the dominant role of Spain, Germany, France, and Italy. The hubs and authorities, Pathfinder and Ps cores methods confirm these observations. 

Using the "right" order of the nodes in a matrix representation can reveal the network structure as block patterns in the displayed matrix. The clustering of network nodes based on corrected Salton dissimilarity again shows the dominant role of Spain, Germany, France, and Italy, but also two main clusters of the developed -- less developed countries division. The Balassa normalization (log(measured/expected) visits) matrix shows that most visits within the two main clusters are above expected, while most visits between them are below expected; within the clusters of Balkan countries, Baltic countries, \{SK, CZ, HU\}, \{IS, DK, NO\} visits are much above expected, etc.
\end{abstract}
\medskip

\noindent\textbf{Keywords:} Weighted network, skeleton, hubs and authorities, generalized cores, Pathfinder, matrix display, clustering, Salton index, Balassa normalization.


\section{Erasmus flow network}

Erasmus+ is a European Union (EU) program designed to support education, training, youth, and sport across Europe. Established in 1987, it aims to provide opportunities for individuals to study, train, gain work experience, and volunteer abroad, while also fostering cooperation and innovation in these fields.
Key features of Erasmus+ are (1) mobility opportunities, (2) cooperation projects, 
(3) policy development, and (4) sport initiatives.


Erasmus+ is funded by the EU, with a budget of over €26 billion for the 2021-2027 period, making it one of the largest programs of its kind.
It is open to EU member states, as well as non-EU countries associated with the program. Millions of individuals and thousands of organizations participate annually.

These papers collectively explore Erasmus mobility through network analysis, regional comparisons, and temporal trends, emphasizing institutional and national patterns in student exchanges \cite{shields2013globalization, breznik2015exploring, dabasi2019international, gadar2020multilayer, breznik2020erasmus, restaino2020analysing, gadar2022cooperation, breznik2024analyzing}.

\begin{figure}
\includegraphics[width=0.5\textwidth]{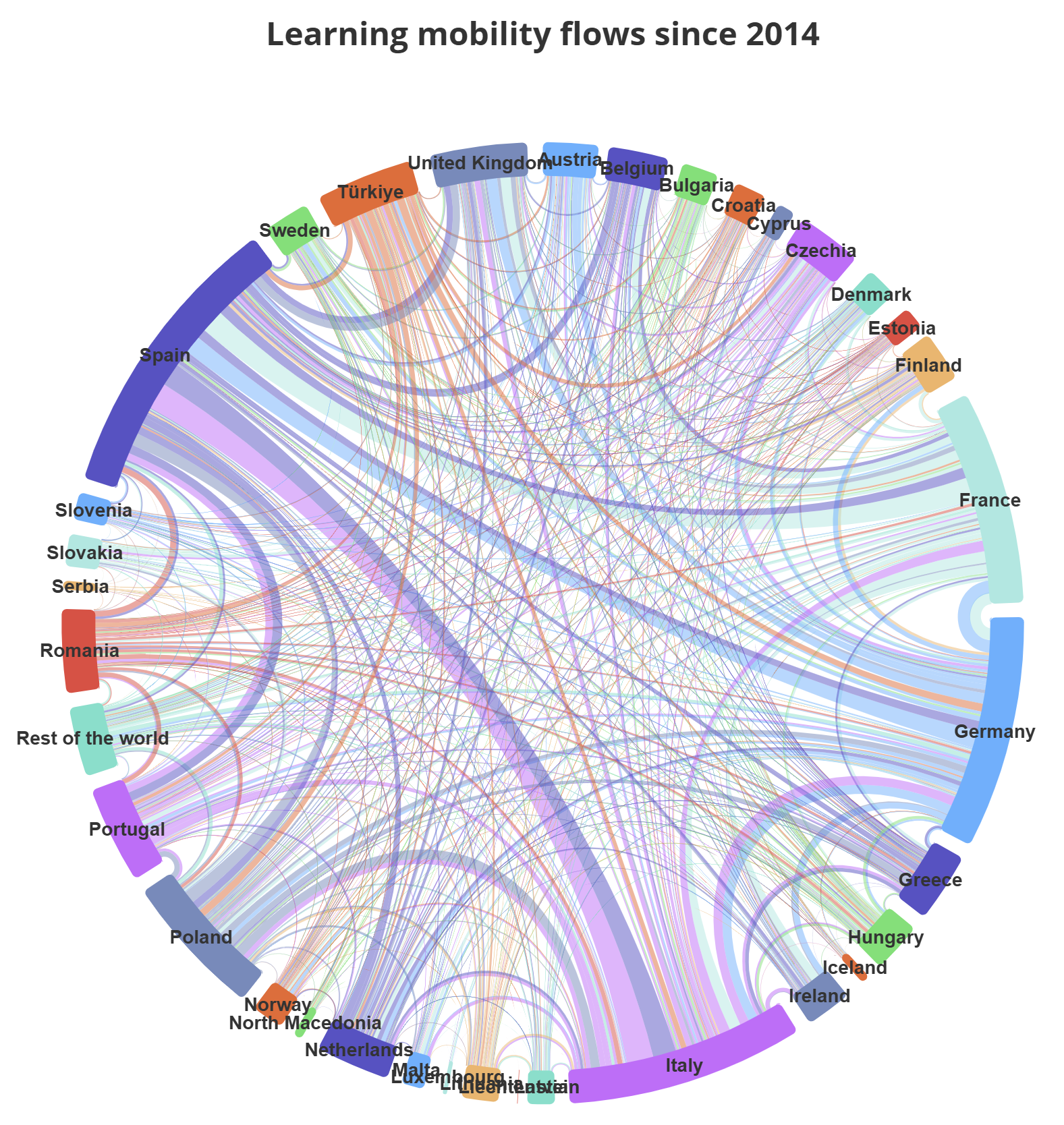}\  \includegraphics[width=0.49\textwidth]{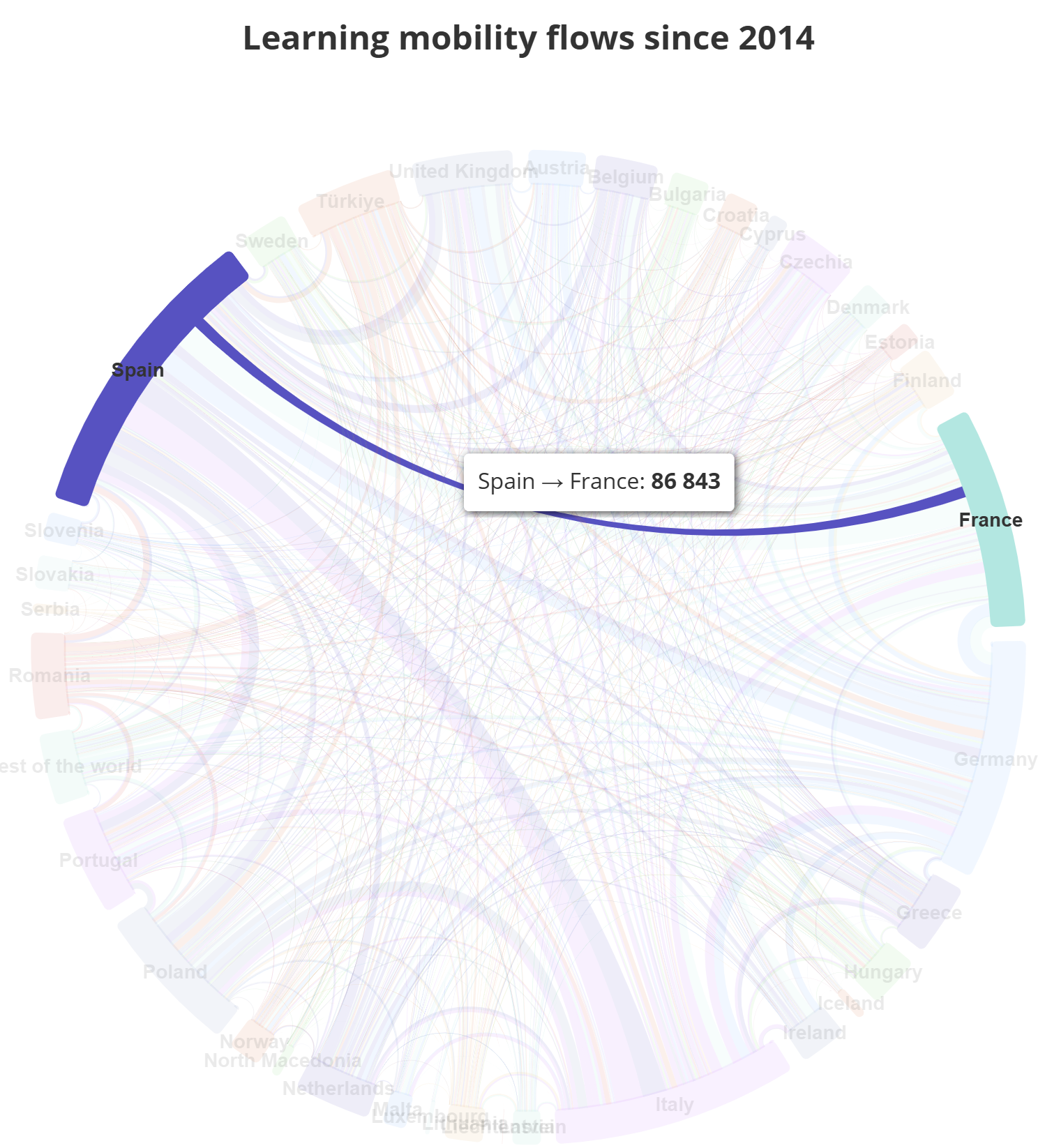}
\caption{Erasmus interactive chart}\label{chart}
\end{figure}

At the bottom of the Erasmus+ page \href{https://erasmus-plus.ec.europa.eu/resources-and-tools/factsheets-statistics-evaluations/statistics/data/learning-mobility-projects}{\textit{Data visualization on learning mobility projects}}, the “Learning mobility flows since 2014” chart can be found -- see the left side of Figure~\ref{chart}.
The interactive chart shows mobility flows between countries since 2014. The colors are related to the sending country. For example, moving the mouse over Italy will highlight all its in/outbound flows and the total count of participants. The same can be done at the flow level -- see the right side of Figure~\ref{chart}.
The interactive chart also provides an option to download the network data -- the network analyzed in this paper.

\subsection{Erasmus mobility network}

We saved the network data in the file \texttt{Learning-mobility-flows-since-2014.csv}. The dataset contains the following countries: Austria (AT),
 Belgium (BE),
 Bulgaria (BG),
 Croatia (HR),
 Cyprus (CY),
 Czechia (CZ),
 Denmark (DK),
 Estonia (EE),
 Finland (FI),
 France (FR),
 Germany (DE),
 Greece (GR),
 Hungary (HU),
 Iceland (IS),
 Ireland (IE),
 Italy (IT),
 Latvia (LV),
 Liechtenstein (LI),
 Lithuania (LT),
 Luxembourg (LU),
 Malta (MT),
 Netherlands (NL),
 North Macedonia (MK),
 Norway (NO),
 Poland (PL),
 Portugal (PT),
 Rest of the world (rW),
 Romania (RO),
 Serbia (RS),
 Slovakia (SK),
 Slovenia (SI),
 Spain (ES),
 Sweden (SE),
 Türkiye (TR),
 United Kingdom (GB).

We used the \href{https://chat.deepseek.com/}{Deepseek} to obtain the corresponding  \href{https://en.wikipedia.org/wiki/ISO_3166-1_alpha-2}{ISO 3166-1 alpha-2} country codes and the total population estimate for each country. Afterward, we converted the collected data into Pajek files \texttt{ErasmusFlows.net}, \texttt{ErasmusFlowsISO.nam}, and \texttt{PopTotal.vec}. The created Pajek files are available at \href{https://github.com/bavla/wNets/blob/main/Data/README.md#Erasmus14}{GitHub/Vlado}.


\subsection{Basic characteristics of the Erasmus mobility network}

A \keyw{network} $\network{N}=(\vertices{V},\edges{L},\functions{P},\functions{W})$
consists of a \keyw{graph} $\graph{G}=(\vertices{V},\edges{L})$, where $\vertices{V}$ is the set of \keyw{nodes} and $\edges{L}$ is the set of \keyw{links} that can be split into two disjoint subsets $\edges{L}=\edges{E}\cup\edges{A}$ -- the set of \keyw{arcs} (directed links) $\edges{A}$ and the set of \keyw{edges} (undirected links) $\edges{E}$.  We denote  $n=|\vertices{V}|$ (number of nodes) and $m=|\vertices{L}|$ (number of links). $\functions{P}$ is a set of node value functions or \keyw{properties} and $\functions{W}$ is a set of link value functions or \keyw{weights}. 

In the Erasmus mobility network, the set of nodes $\vertices{V}$ consists of $n = 35$ countries. The network is directed, $\edges{E}=\emptyset$. There is an arc $(u,v) \in \edges{L}$ from country $u$ to country $v$ iff some persons from country $u$ visited country $v$ under the Erasmus program in the period 2014--2024. Its weight $w(u,v)$ counts the number of such visits. $w \in \functions{W}$. We have two node properties in $\functions{P}$: the function $tp$  that assigns to each country its total population, and the function $iso2$  that assigns to each country its ISO 3166-1 alpha-2 country code.

A standard approach to get insight into the network structure is to draw it. It turns out that it is not so easy. Larger, $n > 20$, dense graphs can't be presented readably with a graphical layout. For the Erasmus network, the density $\gamma = 0.9984$ -- almost all pairs of countries are linked, only two pairs are missing. An alternative visualization is the matrix display of the network.

The second problem comes from weights. They can be represented by link thickness or by the level of grey of matrix cells. The issue is an extensive range and the distribution of weights -- most weights give very thin (invisible) lines or almost white cells. For Erasmus network we have $w_{\min} = 1$ and $w_{\max} = 217003$.


\subsection{Transformations of weights and weight distributions}

\begin{figure}
\includegraphics[width=0.33\textwidth]{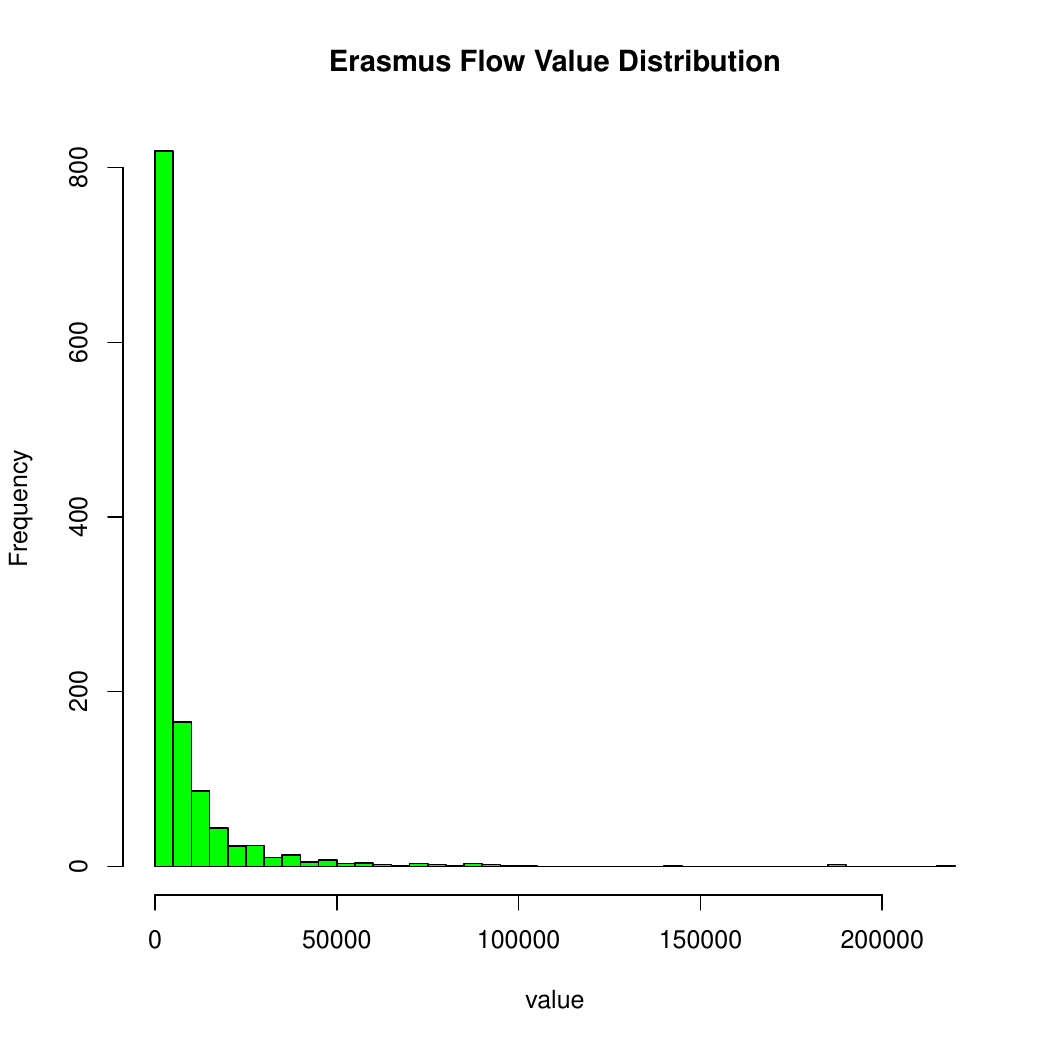}
\includegraphics[width=0.33\textwidth]{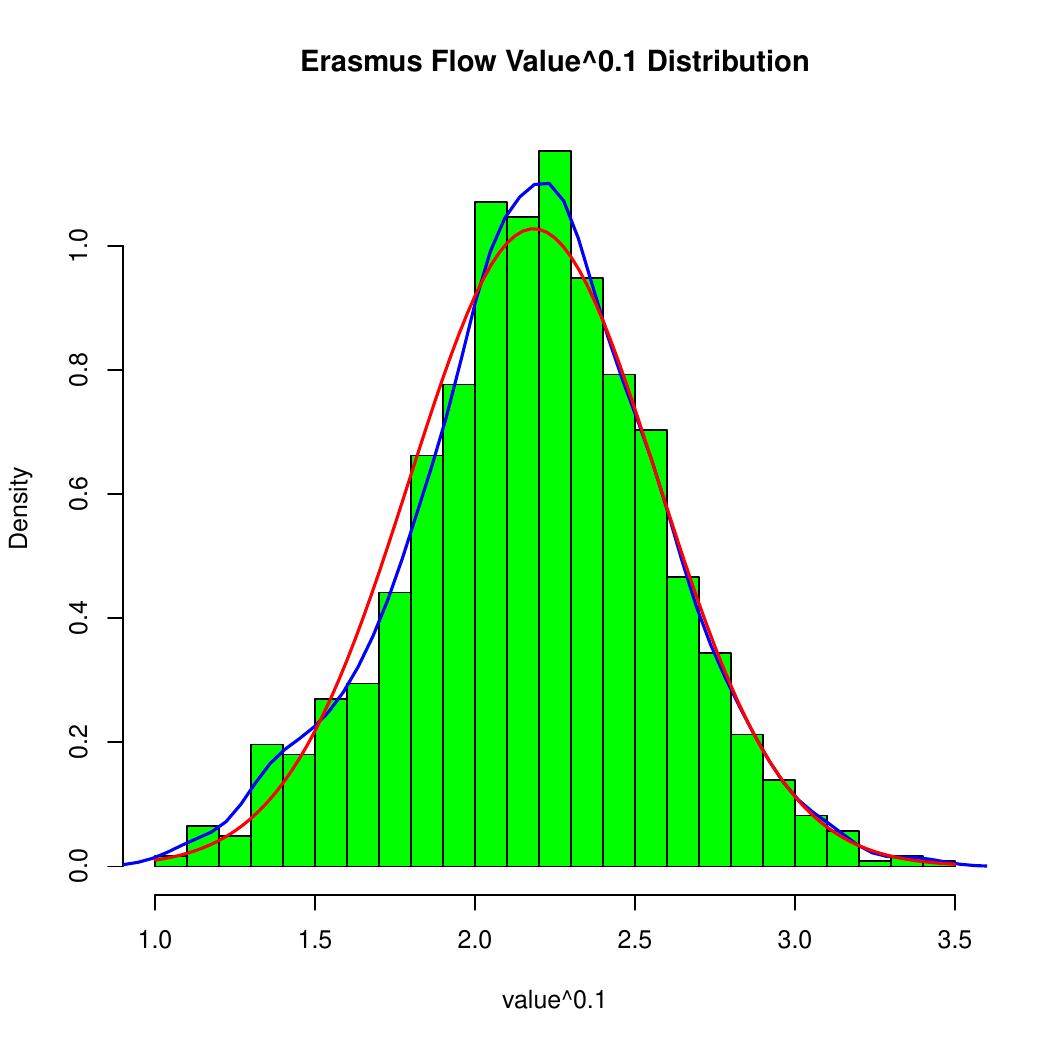}
\includegraphics[width=0.33\textwidth]{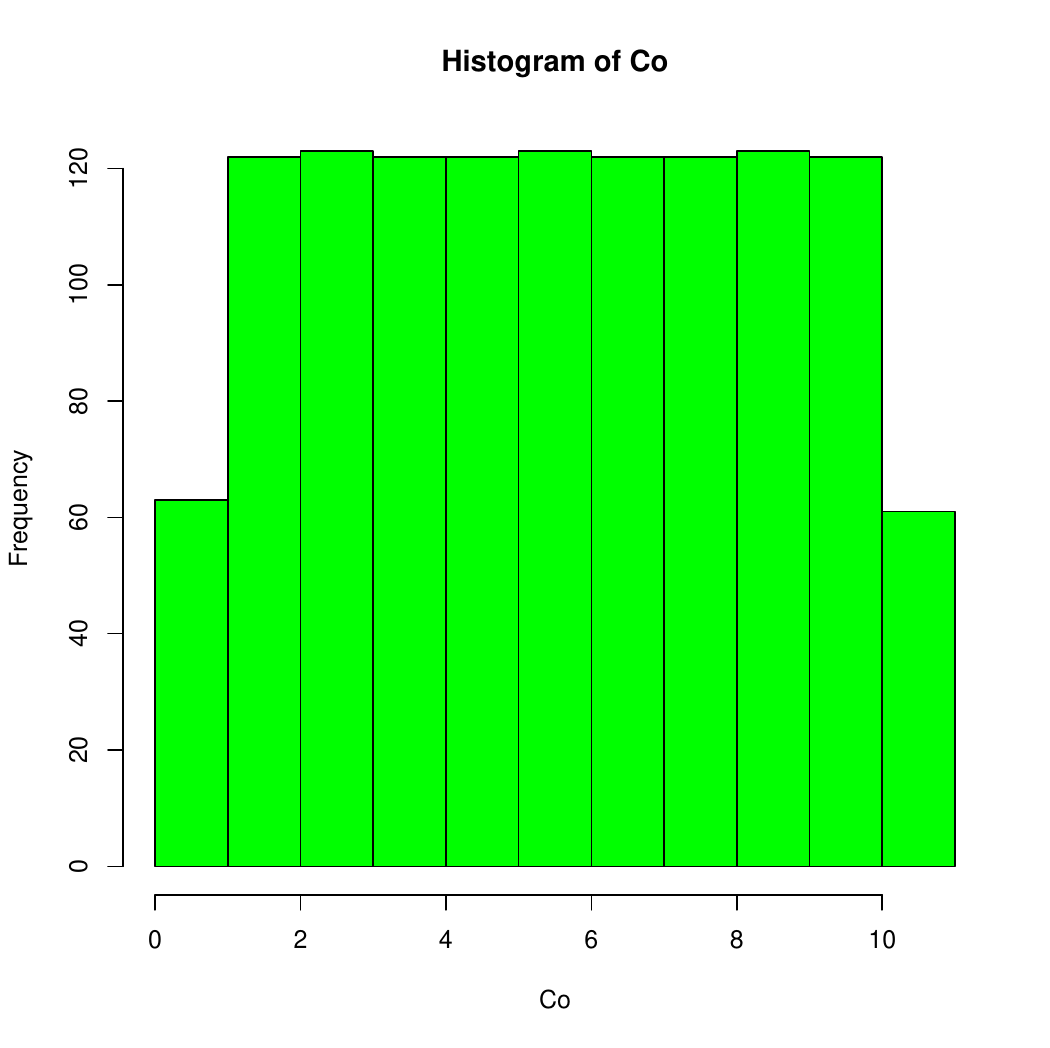}\medskip
\caption{Weight distributions}\label{hist}
\end{figure}

The problem with the large range of weights can be approached by using monotonic transformations.
A function $f : \RR \to \RR$ is a \keyw{non-decreasing transformation} if it has the property
\[  x  < y \Rightarrow f(x) \leq f(y) \]
and is a \keyw{increasing transformation} if it holds
\[  x  < y \Rightarrow f(x) < f(y). \]
Every increasing transformation is also a non-decreasing transformation.
They preserve the order of weights:
Let $w' = f\circ w$ then $w(x)  < w(y) \Rightarrow w'(x) \leq w'(y)$. 

Examples of increasing transformations are $w' = a\cdot w$, $a>0$ or $w' = \sqrt{w}$ or $w' = \log(w)$, etc. In our case, we will use $w' = w^{0.1}$.

The left picture in Figure~\ref{hist} displays the distribution of Erasmus mobility network weights. It shows that most weights are very small and only a small number of links have large weights. After some experiments, we found that the transformation $w' = w^{0.1}$ produces a new weight with almost normal distribution presented in the middle picture in Figure~\ref{hist}. The blue curve is a continuous approximation of the empirical distribution, and the red curve is the corresponding normal distribution. Note that $217003^{0.1} = 3.417013$ and $\ln 217003 = 12.28767$.

In the following, we will use another non-decreasing transformation that splits the weights into intervals with (nearly) equal sizes. We selected 10 intervals. To emphasize the extreme values, we decided to further split one interval into halves and position them at the beginning and the end of the range (see the right picture in Figure~\ref{hist}). The function $w'$ assigns to $x$ the index of the interval to which the value $w(x)$ belongs.

Formally, we can define the function $w'$ as follows. Let $Q_i$, $i \in 1:19$ be the 19-quantiles of the weight $w$ (\cite{WPquantiles}). We additionally set $Q_{20} = \infty$. Then the index of $x$ is $i(x) = \min_{k} (w(x) < Q_k)$ and $w'(x) =  \lceil i(x)/2 \rceil$ where $\lceil x \rceil$ is the ceiling function -- 
the smallest integer that is not smaller than $x$.
 

\subsection{Hubs and authorities}

\begin{table}
\caption{Erasmus hubs and authorities}
\centering
\footnotesize
\renewcommand{\baselinestretch}{0.95}
\begin{tabular}{rlrrrrrrr}
\toprule
 & \multicolumn{2}{c}{\textbf{Country}} & \multicolumn{2}{c}{\textbf{Weighted degree}} & 
   \multicolumn{2}{c}{\textbf{Hubs \& authorities}} & \multicolumn{2}{c}{\textbf{q}} \\
\cmidrule(lr){2-3} \cmidrule(lr){4-5} \cmidrule(lr){6-7} \cmidrule(lr){8-9}
\textbf{Ind} & \textbf{Name} & \textbf{ISO2} & \textbf{Out} & \textbf{In} & \textbf{Hub} & \textbf{Aut} & \textbf{qh} & \textbf{qa} \\
\midrule
1  &           Austria &   AT & 188938 &  217234 & 0.079748 & 0.088240 & 0.948 & 0.999 \\
2  &           Belgium &   BE & 213200 &  240534 & 0.090531 & 0.119336 & 1.158 & 1.005 \\
3  &          Bulgaria &   BG & 169729 &  119420 & 0.062507 & 0.038910 & 0.761 & 0.872 \\
4  &           Croatia &   HR & 116218 &  119898 & 0.041746 & 0.039963 & 0.778 & 0.851 \\
5  &            Cyprus &   CY &  40886 &   84526 & 0.012811 & 0.025387 & 0.701 & 0.742 \\
6  &           Czechia &   CZ & 248912 &  260293 & 0.091028 & 0.096235 & 0.863 & 0.866 \\
7  &           Denmark &   DK & 119418 &  138116 & 0.048440 & 0.056839 & 0.961 & 0.960 \\
8  &           Estonia &   EE &  85311 &   77995 & 0.027829 & 0.026146 & 0.783 & 0.772 \\
9  &           Finland &   FI & 166128 &  219211 & 0.070025 & 0.097640 & 1.040 & 0.998 \\
10 &            France &   FR & 996627 &  627114 & 0.466723 & 0.289367 & 1.077 & 1.109 \\
11 &           Germany &   DE & 973914 &  791268 & 0.411990 & 0.323005 & 0.953 & 1.002 \\
12 &            Greece &   GR & 239679 &  274979 & 0.101686 & 0.110783 & 0.940 & 1.005 \\
13 &           Hungary &   HU & 208243 &  187500 & 0.077653 & 0.068687 & 0.855 & 0.883 \\
14 &           Iceland &   IS &  21817 &   43016 & 0.006826 & 0.015338 & 0.832 & 0.741 \\
15 &           Ireland &   IE &  83818 &  270104 & 0.040001 & 0.152623 & 1.319 & 1.130 \\
16 &             Italy &   IT & 886658 &  896081 & 0.447464 & 0.398546 & 1.038 & 1.195 \\
17 &            Latvia &   LV & 108230 &   86204 & 0.035803 & 0.029629 & 0.802 & 0.783 \\
18 &     Liechtenstein &   LI &   2412 &    2216 & 0.000812 & 0.000708 & 0.745 & 0.797 \\
19 &         Lithuania &   LT & 175171 &  122176 & 0.060994 & 0.043589 & 0.833 & 0.824 \\
20 &        Luxembourg &   LU &  13789 &   28702 & 0.005164 & 0.013391 & 1.089 & 0.887 \\
21 &             Malta &   MT &  24323 &  159215 & 0.008673 & 0.080012 & 1.173 & 0.844 \\
22 &       Netherlands &   NL & 305569 &  269472 & 0.139211 & 0.114219 & 0.989 & 1.079 \\
23 &   North Macedonia &   MK &  52852 &   34191 & 0.014676 & 0.009797 & 0.669 & 0.658 \\
24 &            Norway &   NO &  92329 &  137759 & 0.038256 & 0.061938 & 1.050 & 0.981 \\
25 &            Poland &   PL & 608085 &  468951 & 0.272041 & 0.188551 & 0.939 & 1.059 \\
26 &          Portugal &   PT & 293060 &  460831 & 0.133894 & 0.195632 & 0.991 & 1.082 \\
27 & Rest of the world &   rW & 306823 &  209286 & 0.112267 & 0.076372 & 0.852 & 0.866 \\
28 &           Romania &   RO & 388404 &  250745 & 0.147158 & 0.091176 & 0.849 & 0.897 \\
29 &            Serbia &   RS &  40593 &   27945 & 0.012047 & 0.007687 & 0.642 & 0.703 \\
30 &          Slovakia &   SK & 157337 &  103994 & 0.049428 & 0.033399 & 0.750 & 0.744 \\
31 &          Slovenia &   SI &  98170 &  106327 & 0.034655 & 0.034813 & 0.764 & 0.836 \\
32 &             Spain &   ES & 918245 & 1291788 & 0.382854 & 0.612039 & 1.106 & 0.987 \\
33 &            Sweden &   SE & 142343 &  196526 & 0.064073 & 0.093770 & 1.114 & 1.066 \\
34 &           Türkiye &   TR & 496051 &  254676 & 0.179312 & 0.090357 & 0.828 & 0.856 \\
35 &    United Kingdom &   GB & 261499 &  466488 & 0.136135 & 0.236308 & 1.182 & 1.233 \\
\bottomrule
\end{tabular}

\end{table}

To each node $v$ of a network  $\Net = (\vertices{V},\edges{L},w)$
we assign two values: quality of its content (\keyw{authority})
$x_v$ and quality of its references (\keyw{hub}) $y_v$  (\cite{Kleinberg}).

Good hubs select a good authority, and a good hub points to good authorities
\[ x_v = \sum_{u:(u,v) \in \edges{L}} w(u,v) y_u \qquad
\mbox{and} \qquad  y_v = \sum_{u:(v,u) \in \edges{L}} w(v,u) x_u \]
Let $\mathbf{W}$ be a matrix of network $\Net$ and $\mathbf{x}$ and $\mathbf{y}$
authority and hub vectors. Then we can write these two relations as
$\mathbf{x} = \mathbf{W}^T \mathbf{y}$ and $\mathbf{y} = \mathbf{W} \mathbf{x}$.

We start with $\mathbf{y} = [1,1,\ldots,1]$ and then compute new
vectors $\mathbf{x}$ and $\mathbf{y}$. After each step, we
normalize both vectors. We repeat this until they stabilize. We can show that this procedure converges.
The limit vector
$\mathbf{x}^*$ is the principal eigenvector of the matrix
$\mathbf{W}^T \mathbf{W}$; and $\mathbf{y}^*$ of matrix $\mathbf{W} \mathbf{W}^T$.

There is a strong correlation between a node's hub/authority value and its size (weighted out/in-degree). To neutralize the influence of size, we consider averages.
For the weighted average authority value of node $v$ we get ($\mbox{wod}(v)$ is the weighted out-degree and $\mbox{wid}(v)$ is the weighted in-degree of node $v$)
$$  \bar{x}_v = 
    \frac{\sum_{u:(v,u) \in \edges{L}} w(v,u) x_u}{\sum_{u:(v,u) \in \edges{L}} w(v,u)} = 
    \frac{y_v}{\mbox{wod}(v)} $$
To normalize it, we will use the network's weighted average authority value
$$ \bar{x} = \frac{\sum_u \mbox{wod}(u) \bar{x}_u}{\sum_u \mbox{wod}(u)} = 
    \frac{\sum_u y_u}{\sum_u \mbox{wod}(u)} $$
Similarly, we define the corresponding hub values $\bar{y}_v$ and $\bar{y}$. Now we can define the \keyw{authorityness} $qa(v)$ and \keyw{hubness} $qh(v)$ of a node $v$
$$  qa(v) = \frac{\bar{x}_v}{\bar{x}} = \frac{W \cdot y_v }{\mbox{wod}(v) \cdot \sum_u y_u}  \qquad \mbox{and}  \qquad qh(v) = \frac{\bar{y}_v}{\bar{y}} = 
\frac{W \cdot x_v }{\mbox{wid}(v) \cdot \sum_u x_u}  $$
where $W = \sum_u \mbox{wid}(u) = \sum_u \mbox{wod}(u) = \sum_{u,v} w(u,v)$ is the total weight of links.

The hubs and authorities vectors for the Erasmus mobility network (with original weights) are presented in  Table~1. Let's look at the top countries for each characteristic:

\begin{enumerate}
\item\textbf{wod:}
  FR (996627),
  DE (973914),
  ES (918245),
  IT (886658),
  PL (608085),
  TR (496051),
  RO (388404),
  rW (306823),
  NL (305569),
  PT (293060).

\item\textbf{wid:}
  ES (1291788),
  IT  (896081),
  DE  (791268),
  FR  (627114),
  PL  (468951),
  GB  (466488),
  PT  (460831),
  GR  (274979),
  IE  (270104),
  NL  (269472).

\item\textbf{hub:}
  FR (0.466723),
  IT (0.447464),
  DE (0.411990),
  ES (0.382854),
  PL (0.272041),
  TR (0.179312),
  RO (0.147158),
  NL (0.139211),
  GB (0.136135),
  PT (0.133894).

\item\textbf{aut:}
  ES (0.612039),
  IT (0.398546),
  DE (0.323005),
  FR (0.289367),
  GB (0.236308),
  PT (0.195632),
  PL (0.188551),
  IE (0.152623),
  BE (0.119336),
  NL (0.114219).

\item\textbf{qh:}
  IE (1.319),
  GB (1.182),
  MT (1.173),
  BE (1.158),
  SE (1.114),
  ES (1.106),
  LU (1.089),
  FR (1.077),
  NO (1.050),
  FI (1.040),
  IT (1.038).

\item\textbf{qa:}
  GB (1.233),
  IT (1.195),
  IE (1.130),
  FR (1.109),
  PT (1.082),
  NL (1.079),
  SE (1.066),
  PL (1.059),
  BE (1.005),
  GR (1.005),
  DE (1.002).
\end{enumerate}

For the weighted degrees, hubs, and authorities, the first four positions are occupied by Spain, France, Italy, and Germany, followed by other large countries: Poland, the United Kingdom, Türkiye, Romania, and the Netherlands, Portugal, etc.

Interestingly, the largest hubness $qh$ values have Ireland, the United Kingdom, Malta, Belgium, Sweden, Spain, Luxembourg, etc. (including some smaller countries: IE, MT, LU). These countries are preferred by visitors from high hub value countries. The largest authorityness $qa$ values have the United Kingdom, Italy, Ireland, France, Portugal, the Netherlands, Sweden, Poland, etc. Visitors from these countries prefer high authority destinations.
 
\section{Skeletons}


To get insight into the structure of a large (or/and) dense network, we can reduce it to its skeleton by removing less important links and/or nodes \cite{skelet}.
\begin{enumerate}
\item Most often, the \keyw{spanning tree}, \keyw{link cut}, or \keyw{node cut} is used. An improved version of cuts is the \keyw{islands} approach.
\item In the closest \keyw{$k$-neighbors} skeleton for each node, only the largest $k$ incident links are preserved. This approach is invariant for increasing transformations -- for the original and the transformed weight, we get the same result.
\item
The \keyw{Pathfinder} algorithm was proposed in the 1980s by Schvaneveldt (\cite{PF88, PF90, PF09}) for simplifying weighted networks, where the weight measures a dissimilarity between nodes.
It is based on Minkowski operation
$ a \Mw{r} b = \sqrt[r]{a^r + b^r} $.
For $r=1$, $r=2$, and $r= \infty$ we get
$a \Mw{1} b = a+ b$, $a \Mw{2} b = \sqrt{a^2 + b^2}$, and $a \Mw{\infty} b = \max(a, b)$.
For a path $\pi = (v_0, v_1, \ldots, v_k)$ of length $k$ we define its weight $w(\pi) = w(v_0,v_1) \Mw{r}
 w(v_1,v_2) \Mw{r}  \ldots \Mw{r} w(v_{k-1},v_k) $.
 
The Pathfinder procedure removes from a given network $\Net$ every link $(u,v)$ with its weight larger than 
the minimum weight of all $u$-$v$ paths of length at most $q$. The resulting simplified network is denoted $\PFnet(\Net,r,q)$.

\item Cores are a very efficient tool to determine the most cohesive (active) subnetworks. The subset of nodes $\cling \subseteq \Nodes$ induces a \keyw{$P_s$ core at level $t$} if for all $v \in \cling$ it holds $\mbox{wdeg}_{\cling}(v) \geq t$, and $\cling$ is the maximum such subset  \cite{cores}. We denote the $P_s$ core at level $t$ by $\cling_t$. The $P_s$ cores procedure assigns to each node $v \in \Nodes$ its \keyw{$P_s$ core number}, which is equal to the largest value $t$ such that $v\in \cling_t$.

The subnetwork induced by a core $\cling$ is not always connected. Cores are nested
\[ t < s \Rightarrow \cling_s \subseteq \cling_t \]
\end{enumerate}

In this paper, we will apply the last three approaches.

\subsection{$k$-neighbors}

\begin{figure}
\centerline{\includegraphics[width=120mm]{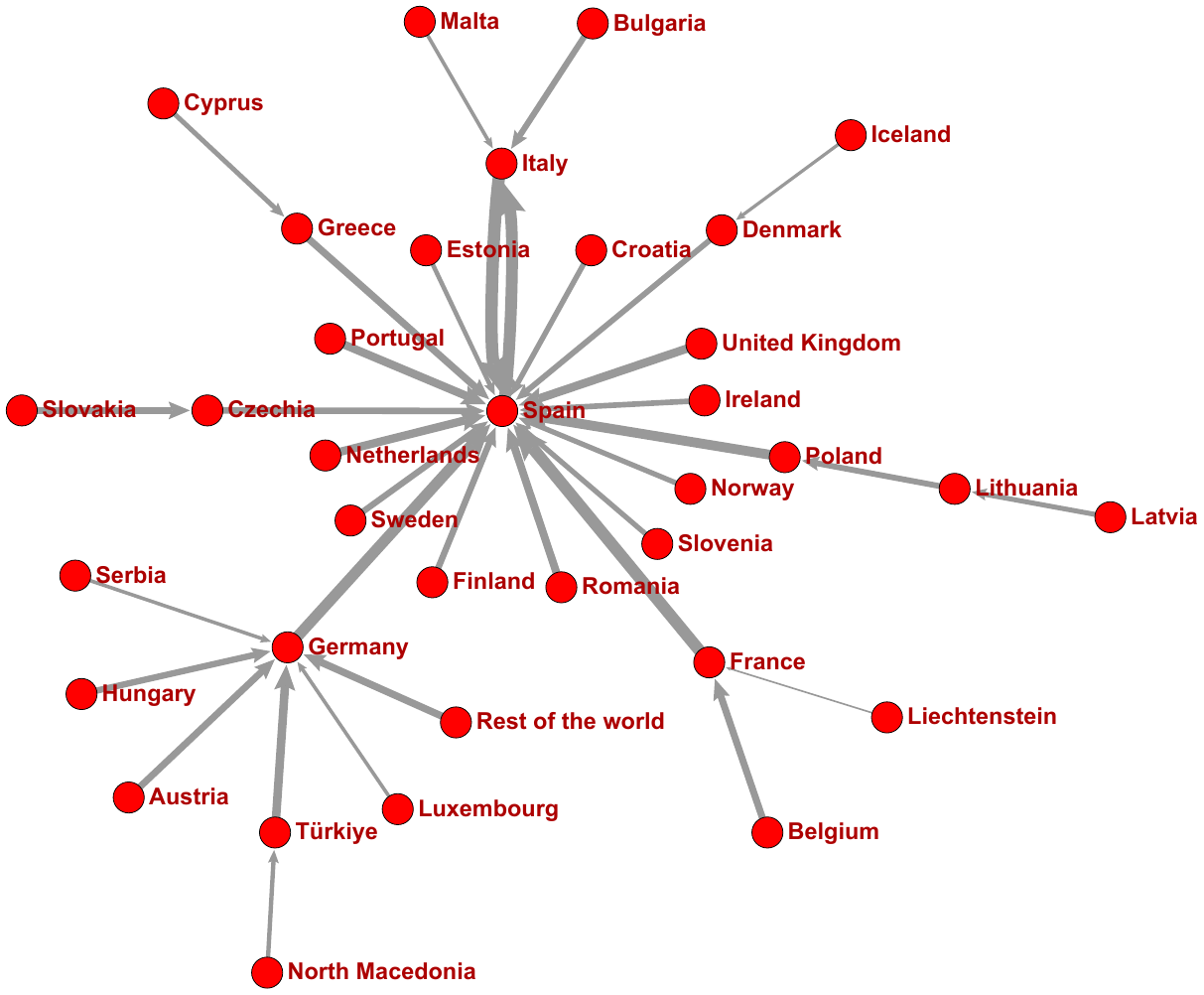}}\smallskip
\caption{Erasmus $1$-neighbors -- first choice\label{one}}
\end{figure}

\begin{figure}
\centerline{\includegraphics[width=120mm]{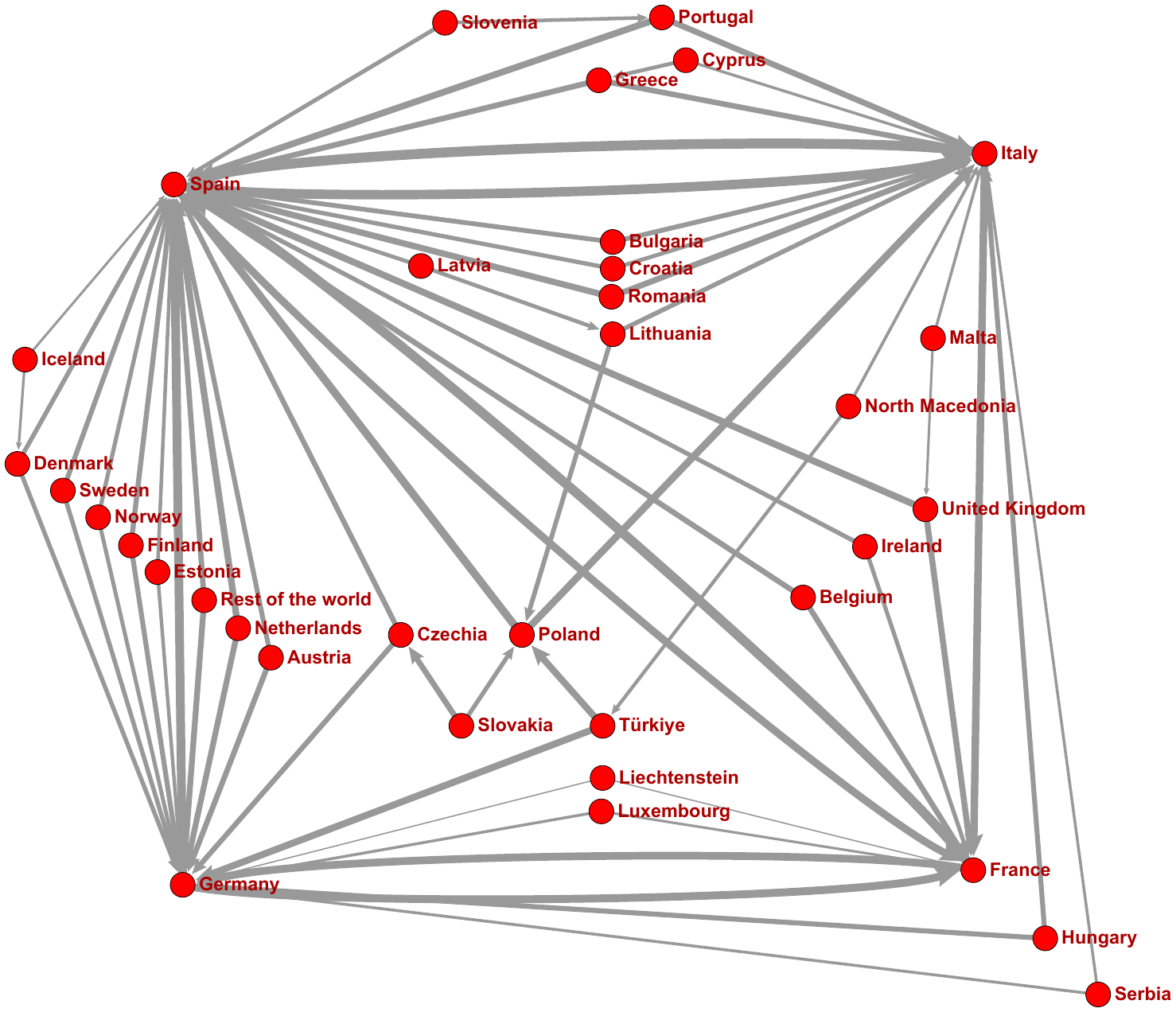}}\smallskip
\caption{Erasmus $2$-neighbors -- first and second choice\label{two}}
\end{figure}

We determined the 1-neighbors and the 2-neighbors skeletons.

\subsubsection{1-neighbors}
In general, connected components of the 1-neighbors skeleton are tree-like directed subgraphs with a single central cycle reachable by a unique path from each other (not on the cycle) component's node.
The Erasmus 1-neighbors skeleton is presented in Figure~\ref{one}.  It highlights Spain as the main attractor in the network and Germany, Italy, France, Poland, Denmark, Greece, and the Czech Republic as secondary attractors.

\subsubsection{2-neighbors}
The Erasmus 2-neighbors skeleton is presented in Figure~\ref{two}. It shows the dominant role of Spain, Germany, France, and Italy. There is a strong mutual interest between Germany and France, France and Spain, and Spain and Italy. The strong interest of the Germans in Spain and the Italians in France is not reciprocal. Visitors from Scandinavian countries, the Rest of the world, Estonia, the Netherlands, the Czech Republic, and Austria prefer Spain and Germany. Visitors from Liechtenstein and Luxembourg prefer Germany and France. Visitors from Hungary and Serbia prefer Germany and Italy. Visitors from Portugal, Greece, Bulgaria, Croatia, Romania, and Poland prefer Spain and Italy. Visitors from the United Kingdom, Ireland, and Belgium prefer Spain and France. Visitors from Slovenia, Cyprus, Malta, North Macedonia, Türkiye, Iceland, Latvia, and Lithuania. Only visitors from Slovakia do not prefer any of the four central countries.


\subsection{$P_s$-cores}

\begin{figure}
\begin{minipage}{0.55\textwidth}
\footnotesize
\renewcommand{\baselinestretch}{0.95}
\begin{tabular}{rlrlrlr}
\toprule
 & \multicolumn{2}{c}{\textbf{All}} & \multicolumn{2}{c}{\textbf{Input}} & \multicolumn{2}{c}{\textbf{Output}} \\
\cmidrule(lr){2-3} \cmidrule(lr){4-5} \cmidrule(lr){6-7}
\textbf{Rank} & \textbf{Id} & \textbf{Value} & \textbf{Id} & \textbf{Value} & \textbf{Id} & \textbf{Value} \\
\midrule
    1  &  DE & 609063  &  DE & 287693  &  DE & 364594 \\
    2  &  FR & 609063  &  FR & 287693  &  FR & 364594 \\
    3  &  IT & 609063  &  IT & 287693  &  IT & 364594 \\
    4  &  ES & 609063  &  ES & 287693  &  ES & 364594 \\
    5  &  PL & 452314  &  GB & 274340  &  PL & 294156 \\
    6  &  GB & 439822  &  PT & 229822  &  TR & 248328 \\
    7  &  PT & 400014  &  PL & 229822  &  RO & 207249 \\
    8  &  RO & 379701  &  IE & 200266  &  rW & 198970 \\
    9  &  TR & 379701  &  RO & 176038  &  PT & 191225 \\
   10  &  GR & 353090  &  CZ & 176038  &  NL & 191225 \\
   11  &  NL & 353090  &  GR & 176038  &  GB & 191225 \\
   12  &  rW & 339887  &  NL & 176038  &  GR & 174407 \\
   13  &  BE & 336319  &  BE & 176038  &  HU & 159516 \\
   14  &  CZ & 330134  &  TR & 176038  &  CZ & 159516 \\
   15  &  IE & 314423  &  rW & 175804  &  BE & 159516 \\
   16  &  HU & 314423  &  AT & 175804  &  BG & 141731 \\
   17  &  AT & 314423  &  FI & 175804  &  AT & 141526 \\
   18  &  FI & 314423  &  SE & 175804  &  SK & 136878 \\
   19  &  SE & 295197  &  HU & 159244  &  LT & 136878 \\
   20  &  BG & 233448  &  MT & 143246  &  FI & 136050 \\
   21  &  LT & 233448  &  DK & 125031  &  SE & 120105 \\
   22  &  SK & 229052  &  NO & 125031  &  DK & 100006 \\
   23  &  DK & 221538  &  BG & 103421  &  HR &  98028 \\
   24  &  NO & 211331  &  LT & 103421  &  LV &  96748 \\
   25  &  HR & 195283  &  HR & 103421  &  SI &  88877 \\
   26  &  SI & 179996  &  SK &  99455  &  NO &  86535 \\
   27  &  MT & 176232  &  SI &  99187  &  EE &  80157 \\
   28  &  LV & 176232  &  LV &  81938  &  IE &  80157 \\
   29  &  EE & 150575  &  CY &  81600  &  MK &  50478 \\
   30  &  CY & 118367  &  EE &  76830  &  CY &  40446 \\
   31  &  MK &  80685  &  IS &  42888  &  RS &  40232 \\
   32  &  RS &  64736  &  MK &  33208  &  MT &  24158 \\
   33  &  IS &  62144  &  LU &  28600  &  IS &  21770 \\
   34  &  LU &  40258  &  RS &  27942  &  LU &  13761 \\
   35  &  LI &   4358  &  LI &   2216  &  LI &   2412 \\
\bottomrule
\end{tabular}
\end{minipage}
\parbox{0.45\textwidth}{
\includegraphics[width=0.45\textwidth,bb=5 90 575 480,clip]{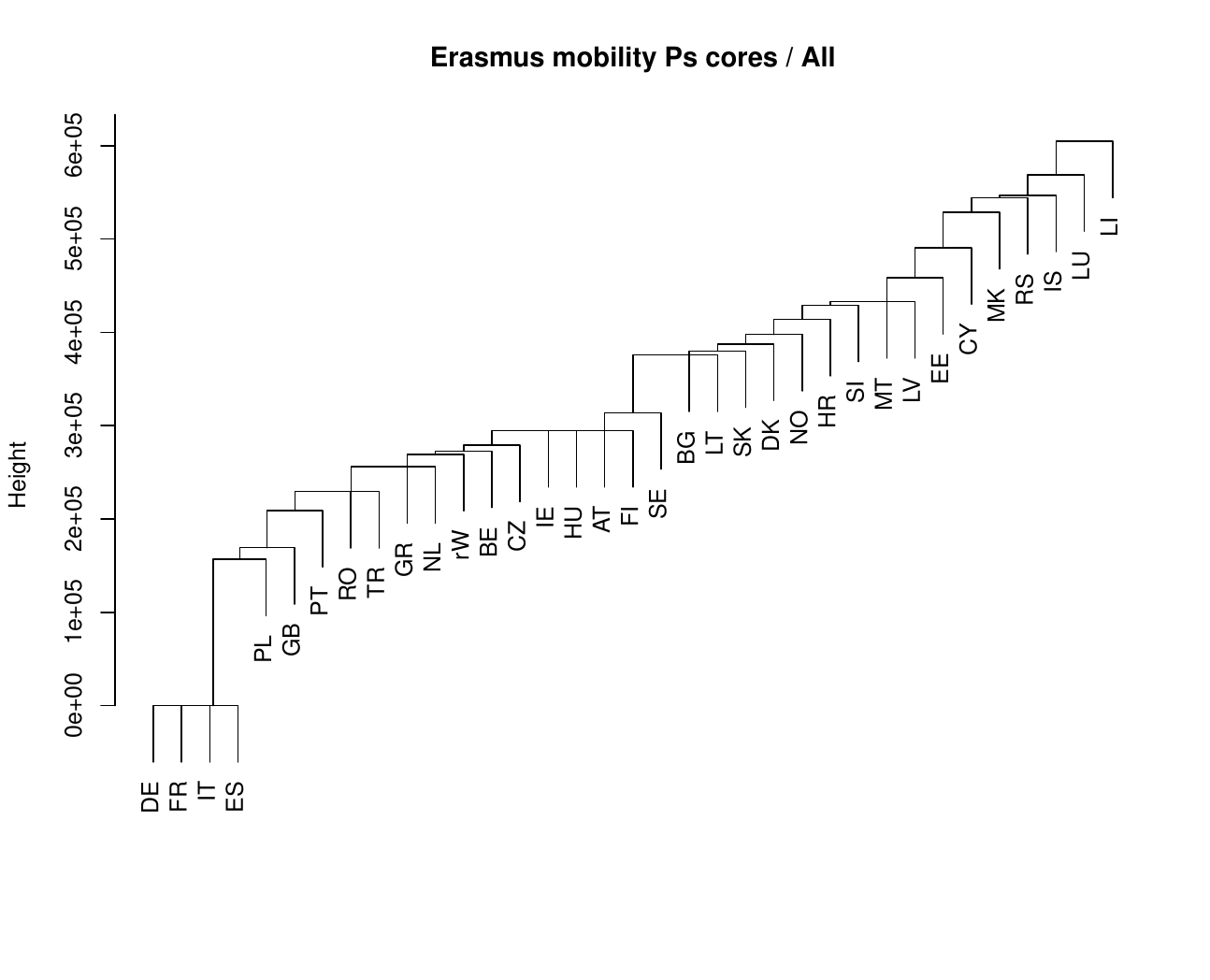}

\includegraphics[width=0.45\textwidth,bb=5 90 575 480,clip]{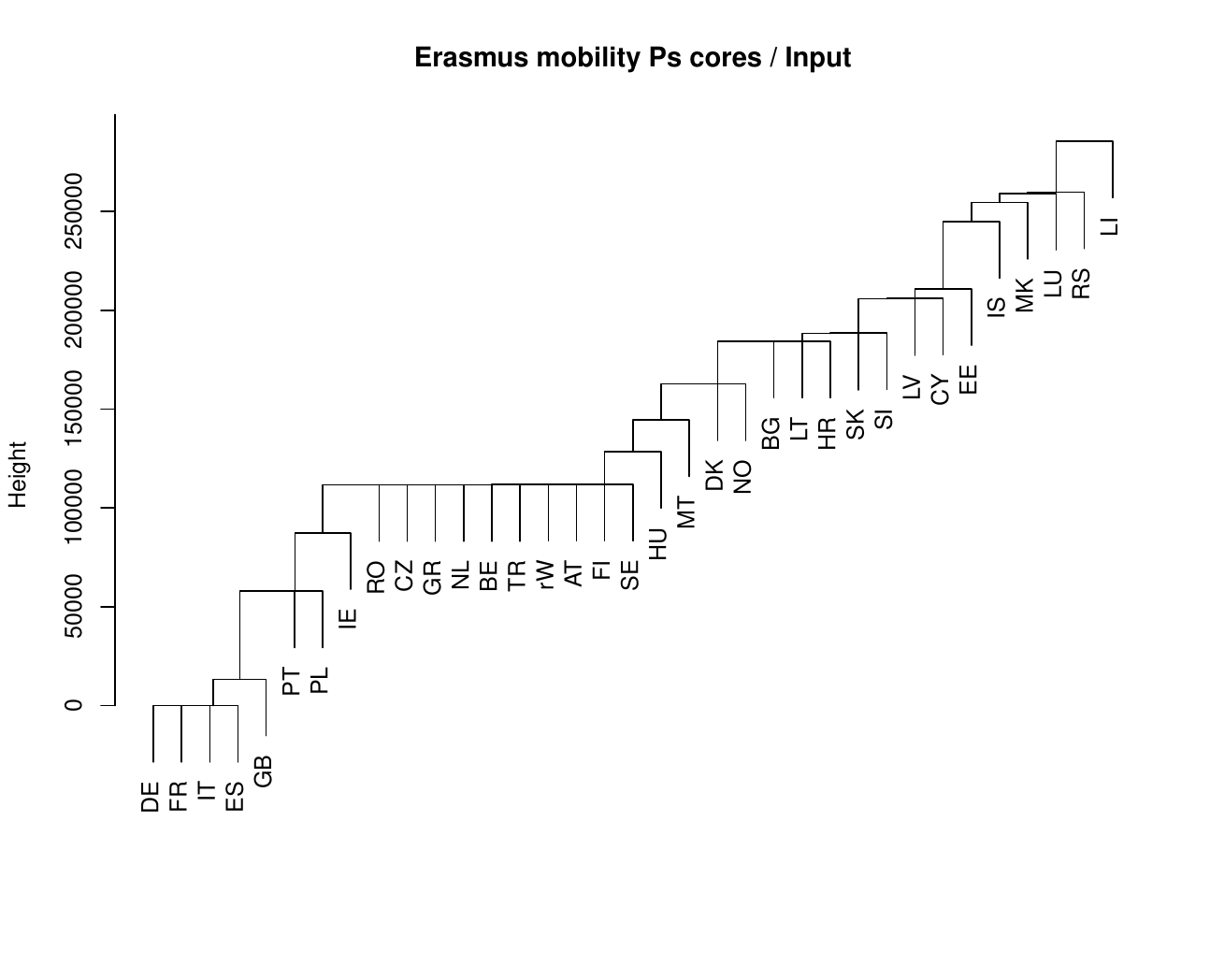}

\includegraphics[width=0.45\textwidth,bb=5 90 575 480,clip]{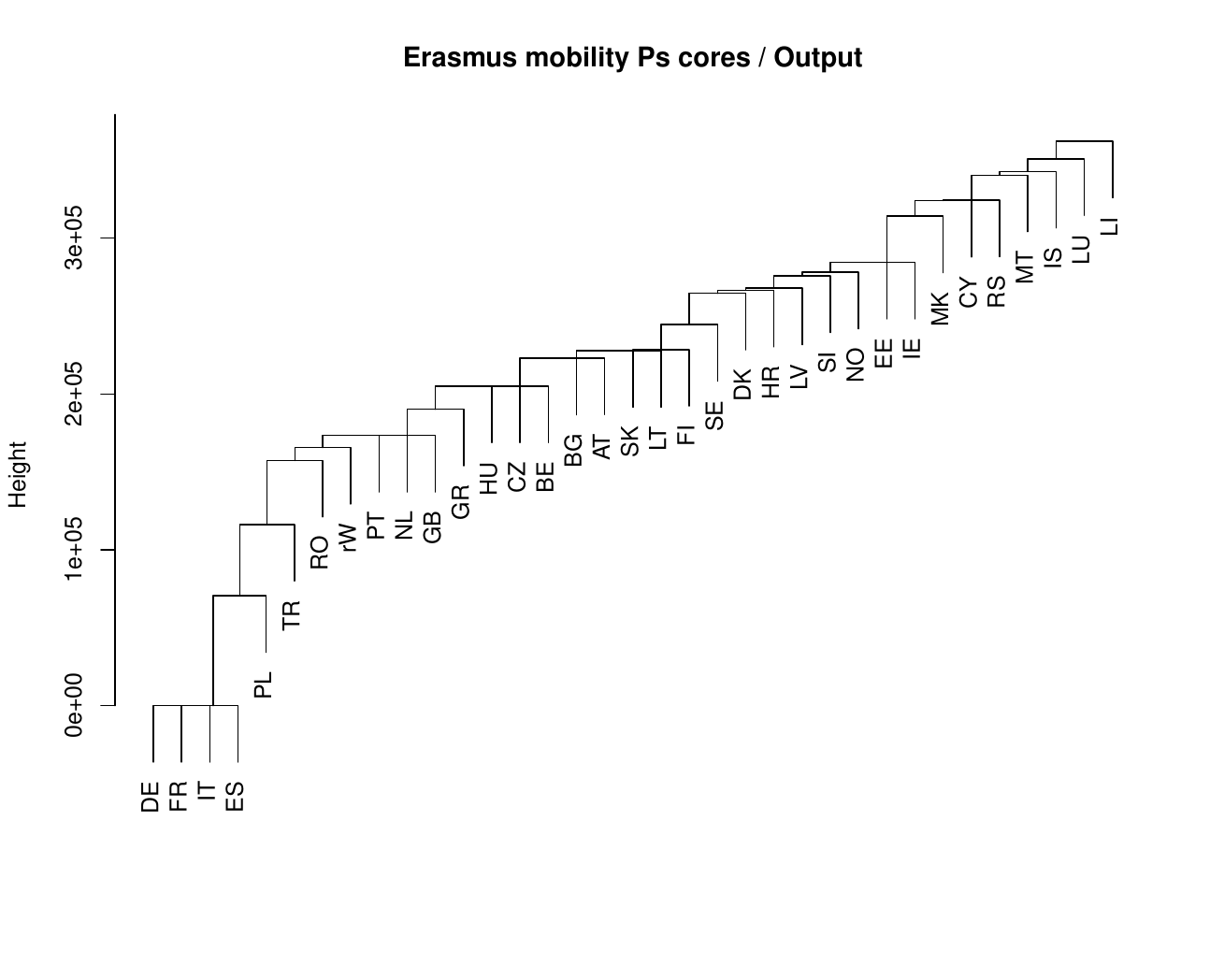}
}
\caption{Erasmus Ps cores}
\label{tab:PsCores}
\end{figure}
These observations are confirmed by the $P_s$ cores approach. We determined $P_s$ core numbers of all nodes (left side of Figure~\ref{tab:PsCores}) for weighted degrees (All), weighted in-degrees (Input), and weighted out-degrees (Output).

We will discuss only the weighted degree cores. The interpretation of the other two results is similar.
The main $P_s$ core is at level 609063 and consists of Germany, France, Italy, and Spain. Each of these four countries exchanged inside the core at least 609063 visitors. At level 452314, they are joined by Poland. The new core is joined at level 439822 by the United Kingdom, and at level 400014 by Portugal. The core is further expanded at the level 379701 by Romania and Türkiye, etc. The expansion process can be visualized using a dendrogram (right side of Figure~\ref{tab:PsCores}). The height in the dendrogram is equal to $t_{\max} - t$.


\subsection{Pathfinder}

The Erasmus network weight $w$ (number of visits) is a similarity measure. The Pathfinder procedure requires a dissimilarity measure $d$. A similarity $w$ can be converted into a dissimilarity $d$ in different ways. For example, $d_1 = w_{\max} - w$ or $d_2 = w_{\max} / w$. We will use the second option.

\begin{figure}
\centerline{\includegraphics[width=120mm]{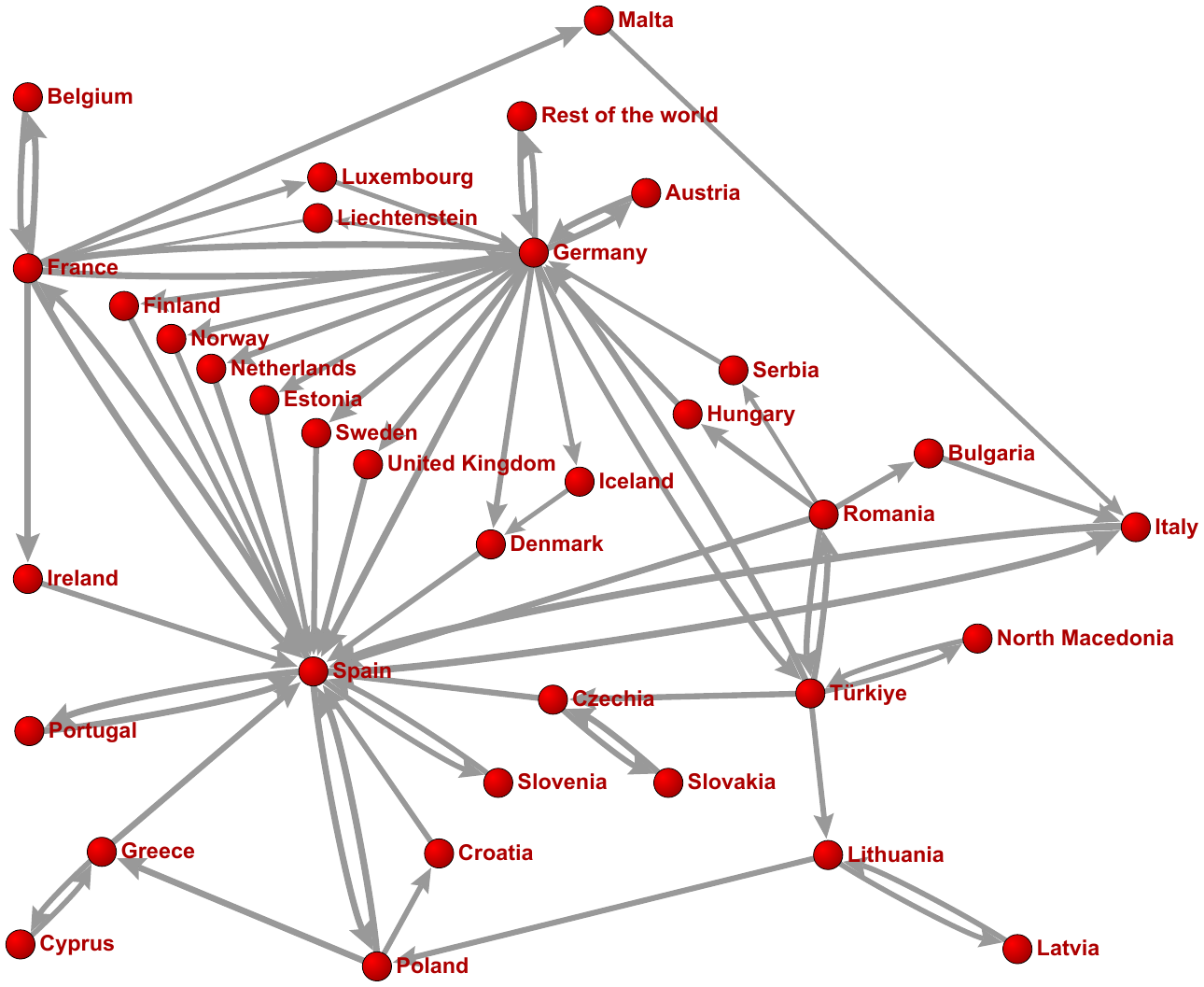}}\smallskip
\caption{Erasmus network Pathfinder skeleton}\label{PF}
\end{figure}

The Erasmus network Pathfinder skeleton in Figure~\ref{PF} is similar to the 2-neighbors skeleton. It emphasizes the central role of Spain and Germany, followed by France, Italy, Türkiye, Romania, and Poland.  Germans prefer France, Finland, Norway, the Netherlands, Estonia, Sweden, the United Kingdom, and Denmark, and they all prefer Spain. There are some expected mutual preferences: Greece and Cyprus, Czechia and Slovakia, Spain and Portugal, Lithuania and Latvia, France and Belgium, Germany and Austria, and Türkiye and North Macedonia.  

\section{Matrix representation}

\begin{figure}
\centerline{\includegraphics[width=100mm]{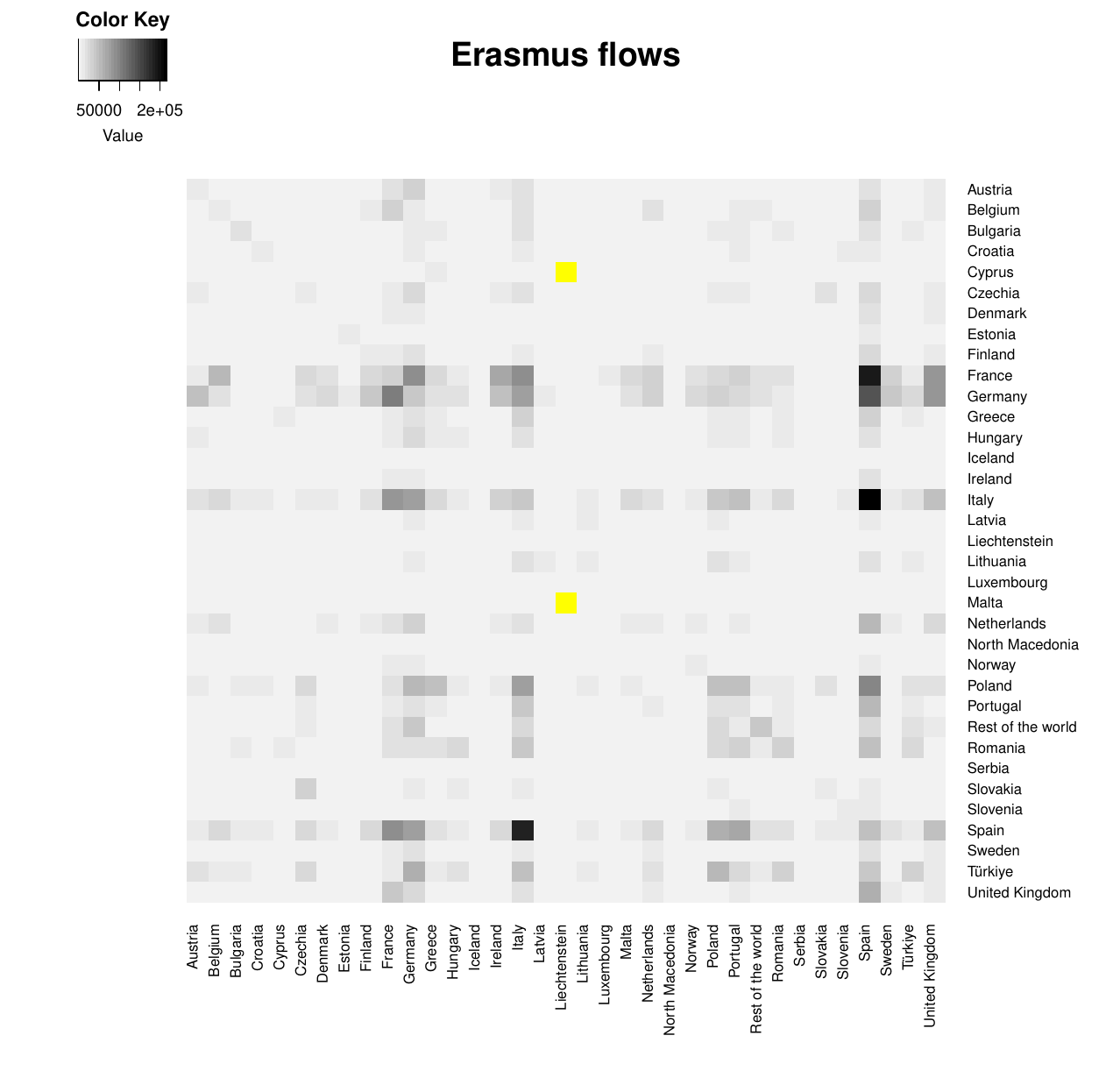}}
\caption{Matrix representation of Erasmus network}\label{matOrg}
\end{figure}

A better option for visualization of dense graphs of moderate size (up to some hundreds of nodes) is the matrix representation. The original Erasmus mobility network matrix is displayed in Figure~\ref{matOrg}.
For each link $(u,v) \in \Links$, its weight $w(u,v)$ is represented in the corresponding square cell with the grey color proportional to its value. Missing links are represented with yellow cells. We see that in the Erasmus network, no visit was made from Cyprus and Malta to Liechtenstein. Most of the cells are light grey -- as expected based on weight distribution (left side of Figure~\ref{hist}). Darker cells belong to large countries, Spain, Italy, Germany, France, the United Kingdom, Poland, Portugal, Romania, and the Rest of the world. The strongest flows are from Italy, France, and Germany into Spain, but there is also a strong flow from Spain into Italy. The readability of the picture could be improved by using increasingly transformed weights.

In the picture, countries are presented in alphabetical order.
A structural order of rows/columns in the matrix representation can be obtained by network clustering \cite{Understand}. Additional reordering of nodes in the hierarchy can be done manually by dendrogram subtrees swapping (left $\leftrightarrow$ right) using R or Pajek.

\subsection{Matrix-based (dis)similarities}

For clustering units (nodes), we need a dissimilarity matrix $\mathbf{D} =[D[u,v]]$ between nodes. In a square weight matrix $\mathbf{W} = [w[u,v]]$, its weights can sometimes be considered (or transformed into) a dissimilarity.
\[ D[u,v] = f(w[u,v],w[v,u]), \quad f(x,y) = f(y,x) \]
For example
\[ D_1[u,v] = \frac{|w[u,v]-w[v,u]|}{\max(w[u,v],w[v,u])} \]
\[ D_2[u,v] = \max(\frac{w[u,v]}{R(u)}, \frac{w[v,u]}{R(v)}), \qquad R(u) = \sum_v w[u,v] \]
Often we use rows (and columns) as node descriptions and apply a selected dissimilarity on them
\[ D[u,v] = d(w[u,.],w[v,.]) \]
where $w[u,.] = [ w[u,1], w[u,2], \ldots, w[u,i], \ldots, w[u,n] ]$ is the row vector of the node $u$.

\noindent
Typical dissimilarities are the \keyw{Euclidean distance}
\[ d_e(\mathbf{x},\mathbf{y}) = \sqrt{(\mathbf{x}-\mathbf{y})^2}\]
and the \keyw{Salton} or \keyw{cosine index}
\[ S(\mathbf{x},\mathbf{y}) = \frac{\mathbf{x} \bullet \mathbf{y}}{\sqrt{\mathbf{x}^2 \cdot \mathbf{y}^2}}, \quad  d_S(\mathbf{x},\mathbf{y}) = 1 - S(\mathbf{x},\mathbf{y})  \quad  \mbox{or}   \quad  d_a(\mathbf{x},\mathbf{y}) = \frac{\arccos S(\mathbf{x},\mathbf{y}) }{\pi}\]
where $\mathbf{x}\bullet\mathbf{y} = \sum_i x_i\cdot y_i$ and $\mathbf{x}^2 = \mathbf{x} \bullet \mathbf{x}$.

\subsubsection{Corrected (dis)similarities}

In traditional (dis)similarities, comparing $w[u,i]$ and $w[v,i]$ we are comparing how $u$ relates to $i$ with how $v$ relates to $i$. The problem arises for $i = u$ and $i = v$. We would need to compare 
$w[u,u]$  with $w[v,v]$  and $w[u,v]$  with $w[v,u]$.
\[ w[u,\cdot] = [ w[u,1],\ldots,w[u,i],\ldots,w[u,u],\ldots,w[u,v],\ldots,w[u,k]] \]
\[ w[v,\cdot] = [ w[v,1],\ldots,w[v,i],\ldots,w[v,u],\ldots,w[v,v],\ldots,w[v,k]] \]
 This leads to \keyw{corrected} (dis)similarities.\medskip

\newpage
\noindent
\keyw{Corrected Euclidean distance}  

\[ d_e'(u, v) = \sqrt{ (w[u, v] - w[v, u])^2 + (w[u, u] - w[v, v])^2 +
\sum_{t \notin \{u, v\}} (w[u, t] - w[v, t])^2 } \]

\noindent
\keyw{Corrected Salton index} 
\[ S'(u,v) = \frac{w[u,.] \bullet w[v,.] + (w[u,u]-w[u,v])\cdot (w[v,v]-w[v,u])}{\sqrt{w[u,.]^2 \cdot w[v,.]^2}} \]
It has the following properties
\begin{enumerate}
\item $S'(u,v) \in [-1,1]$
\item $S'(u,v) = S'(v,u)$
\item $S'(u,u) = 1$
\item $w : \Links \to \RR_0^+ \Rightarrow S'(u,v) \in [0,1]$
\item $S'(\alpha u, \beta v) = S'(u,v), \quad \alpha, \beta > 0$
\item $S'(\alpha u, u) = 1, \quad \alpha > 0$
\end{enumerate}

\subsection{Improved matrix presentation}

\begin{figure}
\includegraphics[width=0.5\textwidth,bb=50 30 560 590,clip]{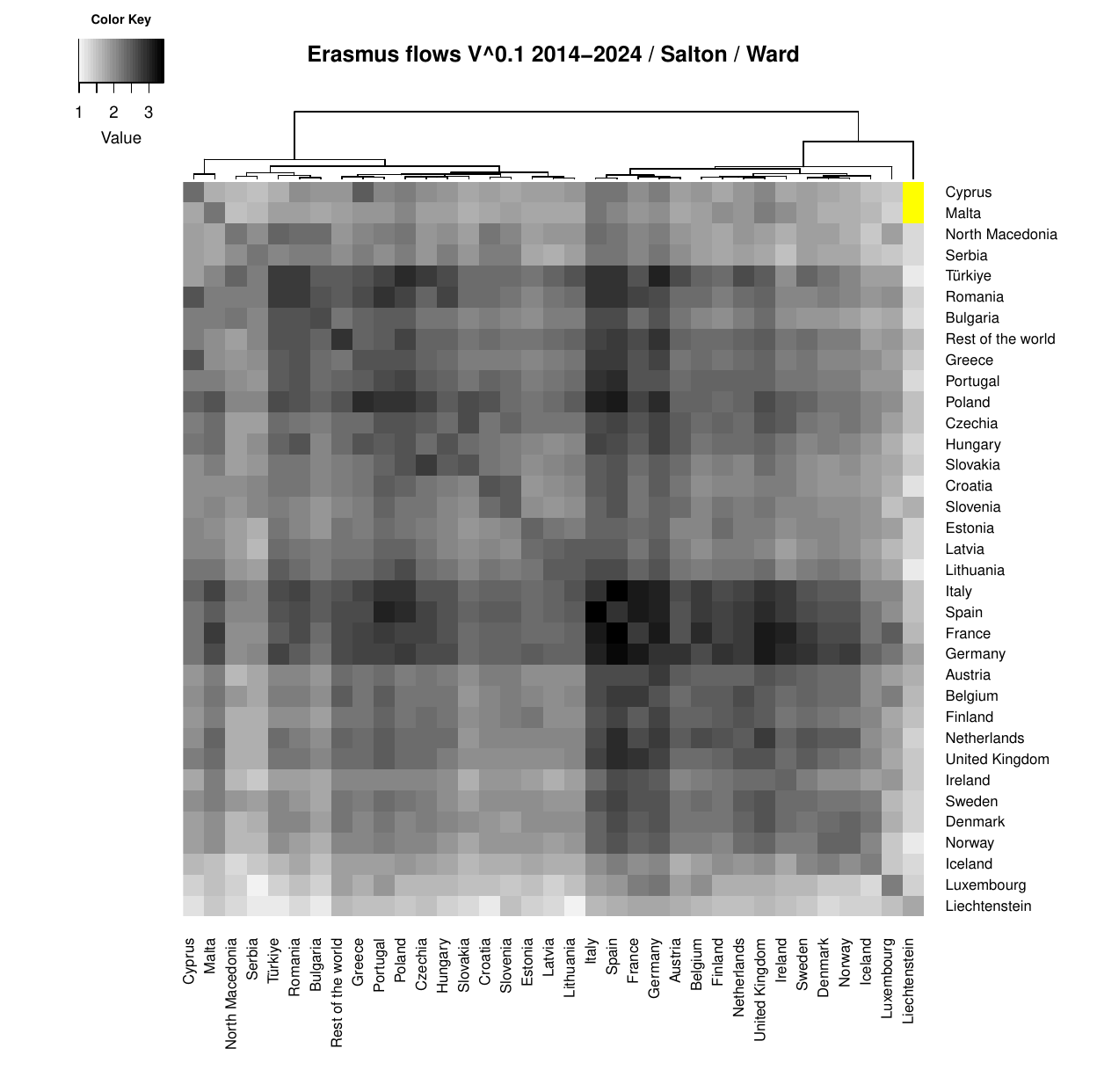}\includegraphics[width=0.5\textwidth,bb=50 30 560 590,clip]{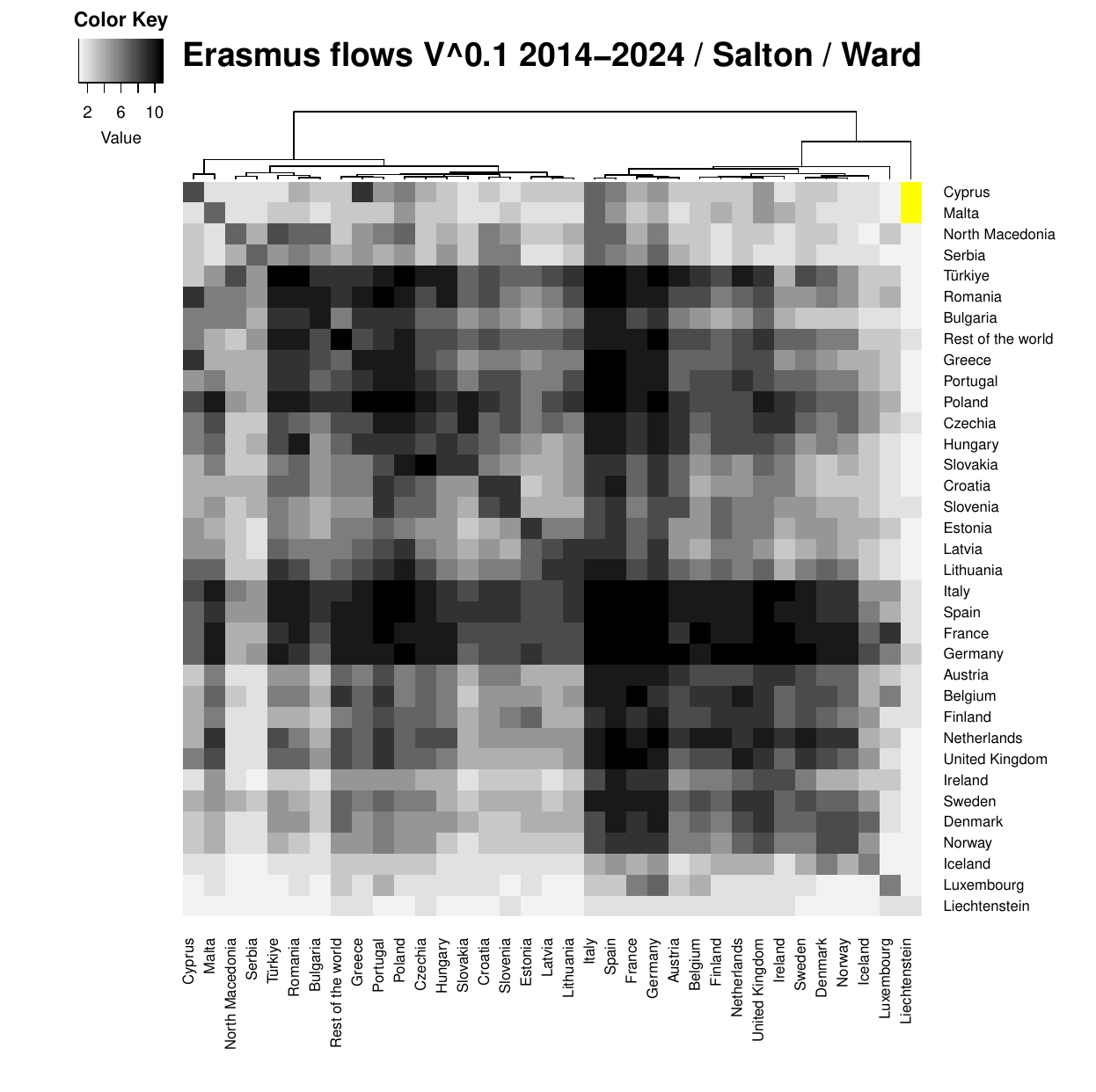}
\caption{Erasmus mobility flow matrix / Salton clustering}\label{Salton}
\end{figure}

To produce a more informative matrix representation, we used the transformed weights $w' = w^{0.1}$, computed the Salton dissimilarity matrix, applied to it the Ward clustering method, and ordered countries according to the obtained dendrogram. We improved the initial picture by some swaps. The final picture is displayed on the left side of Figure~\ref{Salton}.

We can increase the contrast of the produced picture by using (only for visualization) the weight $w'$ based on the 19-quantiles transformation described at the end of Subsection~1.3. The picture is displayed on the right side of Figure~\ref{Salton}.

In the produced picture, we can observe some patterns
\begin{enumerate}
 \item  The ``cross'' formed by the cluster $C_1 = $ (Italy, Spain, France, Germany) -- strong activity with almost all countries in both directions.
 \item Intense diagonal ``squares'' -- clusters: $C_2 = $ (Türkiye, Romania, Bulgaria, Rest of the world, Greece, Portugal, Poland, Czechia, Hungary),  $C_3 = $ (Poland, Czechia, Hungary, Slovakia), $C_4 = $ (Croatia, Slovenia), $C_5 = $ (Estonia, Latvia, Lithuania),
$C_6 = $ (Austria, Belgium, Finland, Netherlands, United Kingdom, Ireland, Sweden, Denmark, Norway), $C_1 \cup C_6$, $C_7 = $ (Sweden, Denmark, Norway, Iceland)
 \item Out-diagonal ``rectangles'': Luxembourg $\times$ (France, Germany),  Greece $\times$ Cyprus, (Croatia, Slovenia)  $\times$ (North Macedonia, Serbia), etc.
 \item In the cross, $C_4 \cup C_5 \cup \mbox{Slovakia}$  less often select France, etc.
\end{enumerate}


\subsection{Normalizations -- activity or Balassa index}

\begin{figure}
\centerline{\includegraphics[width=0.75\textwidth]{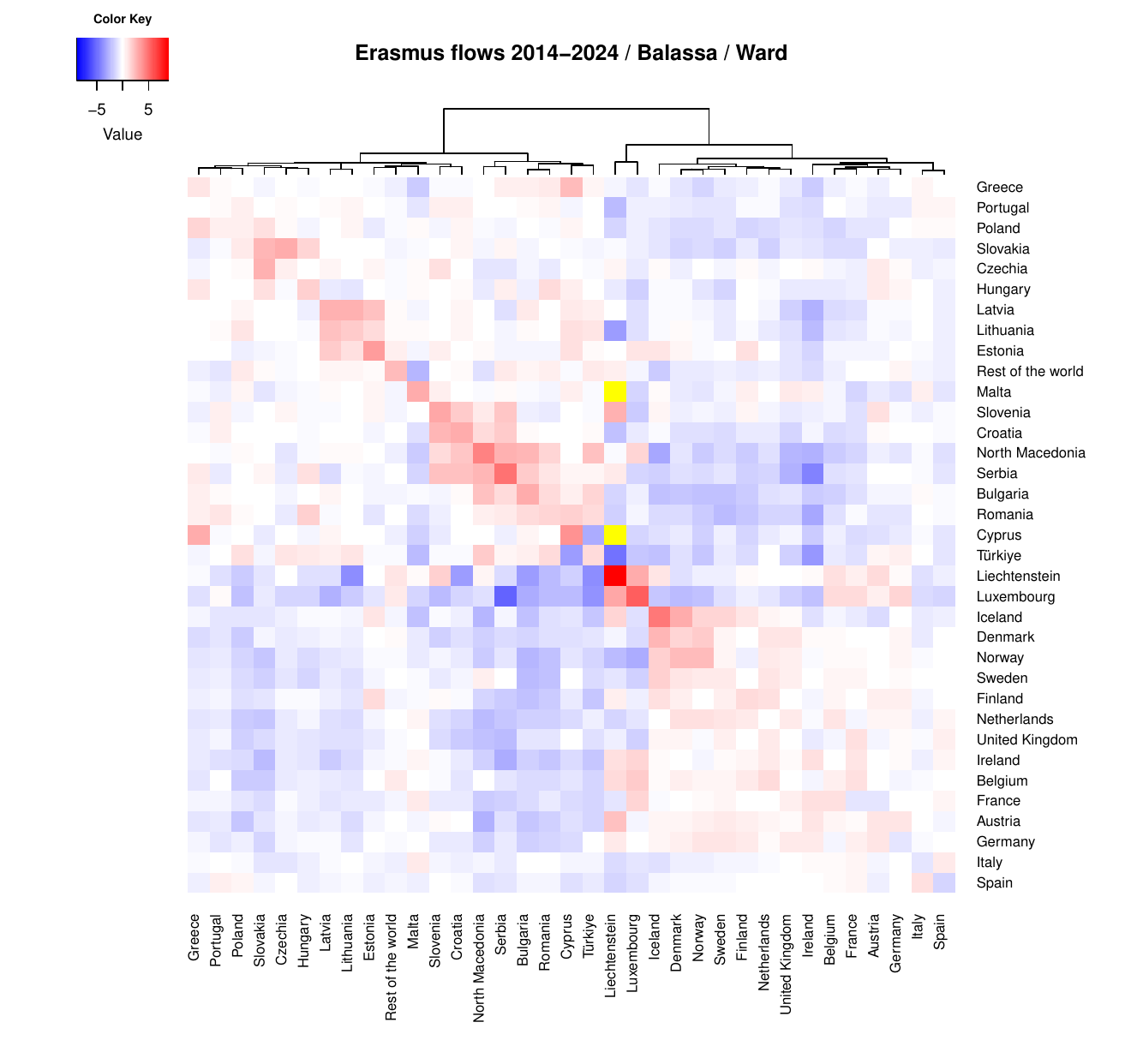}}
\caption{Erasmus mobility flow matrix -- Balassa clustering}\label{Balassa}
\end{figure}

In networks with weights with a large range, usually a few strong nodes prevail. To diminish or neutralize the influence of size on results, different normalizations were proposed and used  \cite{norms,Soviet}. 
In our analysis, we will apply the Balassa normalization.

Let $W = \sum_{e\in \Links} w(e)$.  For $(u,v) \in \Links$ the \keyw{Balassa index} is defined as
\[ A(u,v) = \frac{w[u,v] \cdot W}{\mbox{wod}(u) \cdot \mbox{wid}(v)} \]
and the \keyw{activity normalization} $w'$
\[ w'(u,v) = \log_2 A(u,v) \] 

To produce a matrix representation based on the Balassa approach, after activity normalization, we computed the dissimilarity matrix using the corrected Euclidean distance, applied the Ward clustering method, and ordered countries according to the obtained dendrogram. We improved the initial picture by some swaps. The final picture is displayed in Figure~\ref{Balassa}. The red color represents positive weights, the blue color represents negative weights, and the white color represents the value 0 -- a red cell means that the activity (number of visits) is above expected, a blue cell means that the activity is below expected, and a white cell means that the activity is as expected.

The internal structure of the Balassa matrix is quite interesting

\begin{enumerate}
 \item We notice three main clusters $B_l = $ (Greece : Türkiye) -- less developed, $B_h = $ (Iceland : Spain) -- high developed. and $B_L = $(Liechtenstein, Luxembourg). Most cells inside diagonal squares are red, and out-diagonal rectangles are mostly blue -- exchange between countries from the same cluster is above expected, and below expected between different clusters. 
 \item Red diagonal ``squares'' -- clusters: $B_1 = $ (Slovakia, Czechia, Hungary), $B_2 = $(Latvia, Lithuania, Estonia), $B_3 = $ (Slovenia, Croatia, North Macedonia, Serbia),  $B_4 = $ (North Macedonia, Serbia, Bolgaria, Romania), $B_L$. The exchange between Cyprus and Türkiye is below expected. In the main cluster $B_h$, we can identify a subcluster  $B_5 = $ (Iceland, Denmark, Norway, Sweden, Finland, Netherlands, United Kingdom). Within the clusters $B_1$, $B_2$, $B_3$, $B_4$, $B_L$, and (Iceland, Denmark, Norway), visits are much above expected.
 \item Countries from the cluster $B_l$ are selecting Malta below the expected. The exchange between $B_5$ and $B_L$ is below expected.
 \item Exchange between Cyprus and Greece is above expected.
 \item Exchange of Italy, Spain, and Estonia with other countries is mostly close to as expected.
\end{enumerate}


\subsection{Blockmodeling}

\begin{figure}
\includegraphics[width=0.5\textwidth,bb=30 35 475 505,clip=]{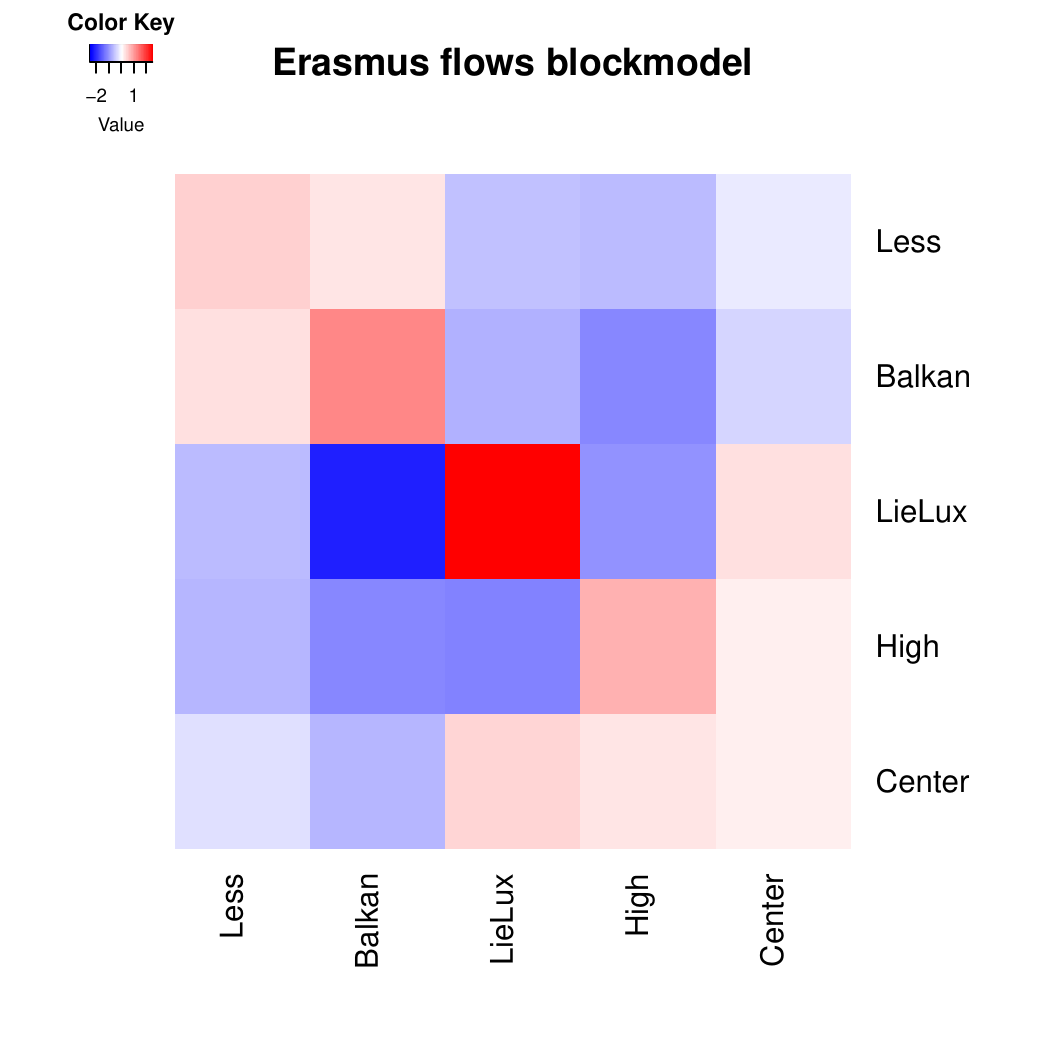}
\raisebox{8mm}{\parbox[b]{0.5\textwidth}{\small
\begin{enumerate}
 \item \textbf{Less:} Greece, Portugal, Poland, Slovakia, Czechia, Hungary, Latvia, Lithuania, Estonia, Rest of the world, Malta,           
 \item \textbf{Balkan:} Slovenia, Croatia, North Macedonia, Serbia, Bulgaria, Romania, Cyprus, Türkiye,       
 \item \textbf{LieLux:} Liechtenstein, Luxembourg,  
 \item \textbf{High:} Iceland, Denmark, Norway, Sweden, Finland, Netherlands, United Kingdom,
 \item \textbf{Center:} Ireland, Belgium, France, Austria, Germany, Italy,  Spain.  
\end{enumerate}\par\vspace*{5mm}}}
\caption{Erasmus mobilty flow blockmodel -- Balassa clustering}\label{BM}
\end{figure}

The main observations about the Balassa matrix can be summarized in a blockmodel displayed in Figure~\ref{BM}. It is based on five clusters \textbf{Less}, \textbf{Balkan}, \textbf{LieLux}, \textbf{High}, and \textbf{Center}.

\begin{enumerate}
 \item The activity inside diagonal clusters \textbf{Less} $\cup$ \textbf{Balkan}, \textbf{LieLux},  \textbf{High} $\cup$ \textbf{Center}, and  \textbf{LieLux} $\cup$ \textbf{Center} is above expected.
 \item There is very strong activity inside the cluster \textbf{LieLux}.
 \item Surprisingly, the mutual activity between clusters  \textbf{LieLux} and \textbf{High} is below expected.
 \item Flow from  \textbf{LieLux} into \textbf{Balkan} is strongly below expected.
 \item Flow from all countries into \textbf{Center} is close to as expected. Activities of \textbf{Center} and \textbf{Less} are closer to as expected -- the central $3\times 3$ square is darker.
\end{enumerate}



\section{Conclusions}

Through an exploratory network analysis of the Erasmus mobility network, we have revealed its basic structure, which leads to some interesting questions for future research, such as: Why is Spain the most attractive country? How to reduce the intensity in the blue rectangles (the gap) between the clusters of less developed and highly developed countries? How do the European Union's Erasmus+ and Horizon programmes affect scientific cooperation between European countries? To address such questions, a temporal version of the Erasmus mobility network and additional data (neighbourhood relations, population size, GDP, etc.) are needed.


Analyzing the Erasmus mobility network, we illustrate typical problems and approaches in the analysis of weighted networks.


\section{Acknowledgments}


The computational work reported in this paper was performed using R and Pajek. The code and data are available at  \href{https://github.com/bavla/wNets/blob/main/Data/README.md#Erasmus14}{GitHub/Vlado/wNets}.

This work is supported in part by the Slovenian Research and Innovation Agency (ARIS research program P1-0294 and research project  J5-4596), and prepared within the framework of the COST action CA21163 (HiTEc).



\end{document}